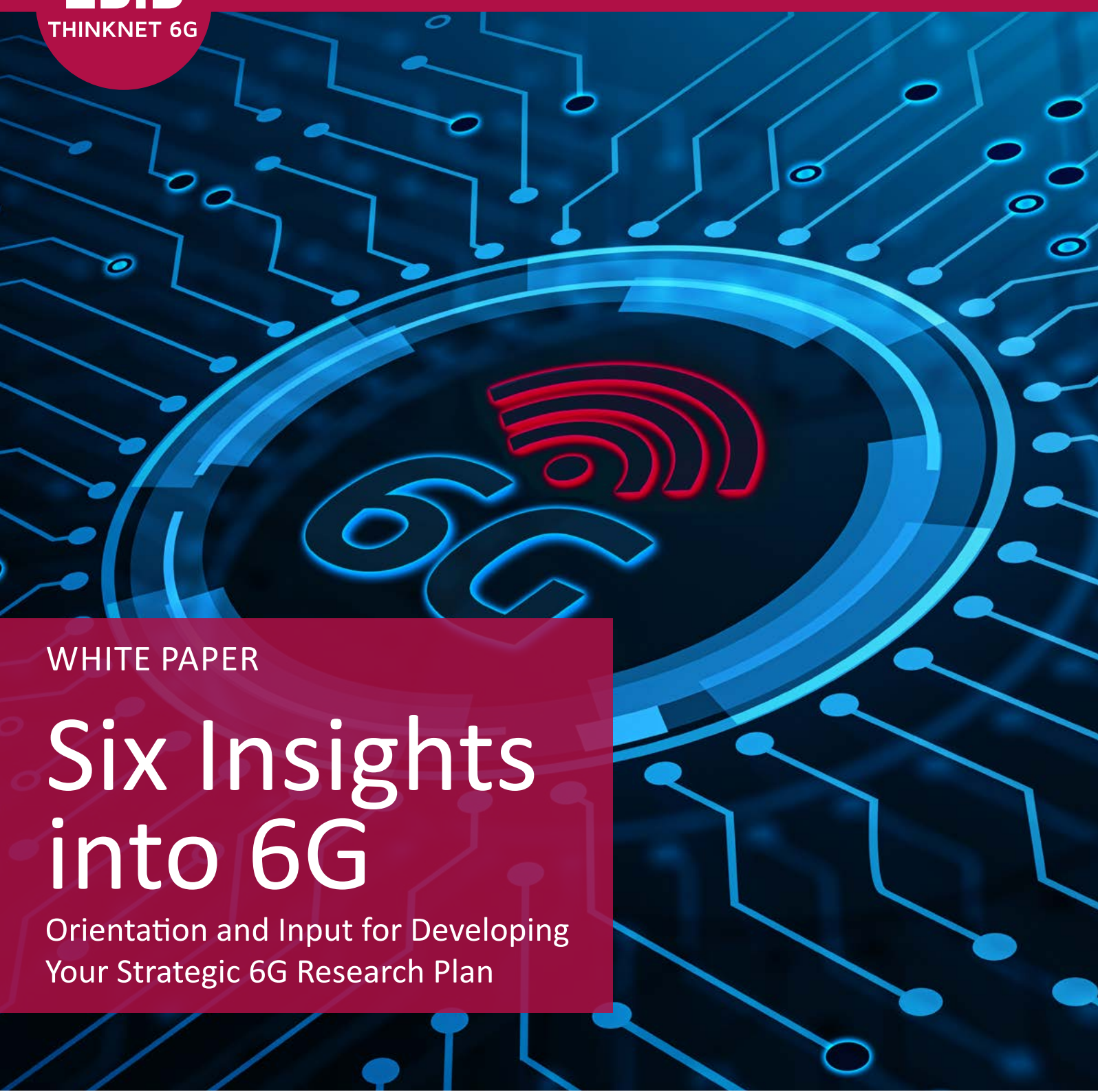

**ZD.B**
THINKNET 6G

bayern innovativ
Innovation leben.

WHITE PAPER

# Six Insights into 6G

Orientation and Input for Developing Your Strategic 6G Research Plan

MÜNCHNER KREIS

# Table of Contents

# Executive Summary

This document is a summary of the findings from a series of workshops which were held by Thinknet 6G [1] and MÜNCHNER KREIS [2] in 2021, with the goal to provide orientation and input for developing a strategic 6G research plan. The topics selected for the workshops are aspects of 6G that we expect will have a significant impact on other industries and on society:

- 6G as both a communication infrastructure and a sensing infrastructure
- The extensive use of artificial intelligence in 6G
- The security and resilience of 6G

This paper does not go into the technical details of how to develop and implement 6G. Rather, it provides input from experts from both the wireless industry as well as from other sectors about (mostly) non-technical topics that will need to be addressed in parallel with the technical developments, such as new use cases, regulation, communication with the public, and cross-industry cooperation.

We have identified six areas that will have a significant impact on the development and use of 6G, and that organizations must consider as they begin their plans and designs for 6G. These are:

### 1. Cross-sectoral, Interdisciplinary and International Research and Development

The full potential of 6G can only be achieved when different industries and domains communicate and cooperate with each other to harness the power of 6G. Industry sectors that will be the main use cases for the sensing infrastructure must be included in the discussion from an early stage, and research and development related to 6G must be coordinated globally to develop the standards that ensure interoperability.

### 2. Communication with the Public and Technology Acceptance

Proactive, open and honest communication with the public about the risks and benefits of both 6G and AI is critically important for technology acceptance. For example, explainable AI can improve acceptance of AI, and methods to ensure security & privacy in a sensing network can improve acceptance of the sensing infrastructure. Opportunities and the 6G roadmap must be communicated realistically, not just evangelized. Involving professional science communicators and sociologists will facilitate understanding and then acceptance. Developing functioning, secure and easy-to-understand use-cases and success stories will demonstrate both the feasibility and value of 6G.

### 3. Security, Privacy, Trust and Resilience

New concepts such as joint communication and sensing will only obtain public acceptance when they are secure, when they protect privacy, and when the public is comfortable placing trust in the systems. Proper security in design and operation creates a unique selling point and can be a geo-political advantage or a mandatory requirement. It is critical to embed security by design and privacy by design into all network devices and all end devices. Zero-trust networking and zero-trust environments will be needed. The more society comes to rely on immersive, embedded communication and sensing, the more important resilience in all its dimensions (availability, reliability, service levels, resistance to attacks, self-healing networks…) becomes.

### 4. Managing Complexity, Orchestration and New Technologies

Using the 6G network for joint communication and sensing necessitates a new level of complexity not yet addressed in 5G. 6G will require fusion at multiple levels in the technology: data fusion, sensor fusion, network fusion, ... And all of this across multiple providers, network infrastructures, ICT (e.g., cloud) infrastructures, and multiple sectors and domains. Standardizing, orchestrating and maintaining control over this level of complexity will be a huge challenge. AI is expected to play a major role here to provide solutions to cope with complexity.

### 5. New Business Models

6G technology will enable new use cases and services that the industry has not yet thought of, or that were not yet technically possible. As the technology advances and new insights are obtained, use cases will become more individualized and services will be optimized for the evolving 6G capabilities. We expect better user experience, not only for humans but also for IoT devices, sensors, actuators, and machines. These new possibilities will spawn new companies and new business models which take advantage of sensing data. In addition, we expect new forms of cooperation between incumbent operators and interdisciplinary cooperation between enterprises from different sectors.

### 6. Standardization and Regulation

6G stakeholders require the surety that regulation and standardization provide in order to develop new products. Both regulation and standardization must be clear, must take place early in the process, and must be agreed worldwide. New in 6G is the requirement to standardize across a significantly increasing number of multiple network technologies, multiple operators, multiple sensor providers and multiple verticals. Numerous areas that require standardization and/or regulation were identified in the workshops. These are listed in the "Recommendations" section below.

Based on these six impact areas and on the discussion in the workshops, we compiled a list of the top 10 recommendations for specific areas where organizations should place their focus when developing their strategic plan for 6G. These are:

1) Interdisciplinary and intersectoral communication
2) International cooperation
3) Develop and enforce standards and regulation
4) Trust and communication with the public
5) Invest in building up and retaining expertise
6) Security-by-design and privacy-by-design
7) Focus on explainable, reliable AI with measurable KPIs
8) Form data alliances and common data pools
9) Managing system complexity
10) Focus on resilience and availability

For our readers who are involved in 6G research, be it at a university, at a research institute or in industrial research, we also included a summary of the top 10 areas that require additional research, again based on the input received in the workshops. The top 10 research areas for 6G we identified are:

1) Joint communication and sensing
2) 6G IoT
3) 6G end-to-end networking
4) Resilient complex systems
5) Service enablement
6) Human-centricity and digital trust
7) Security and privacy
8) AI and ML for wireless networking
9) Data
10) Softwarization

The top 10 Recommendations and top 10 Research Needs are detailed in the respective chapters below.

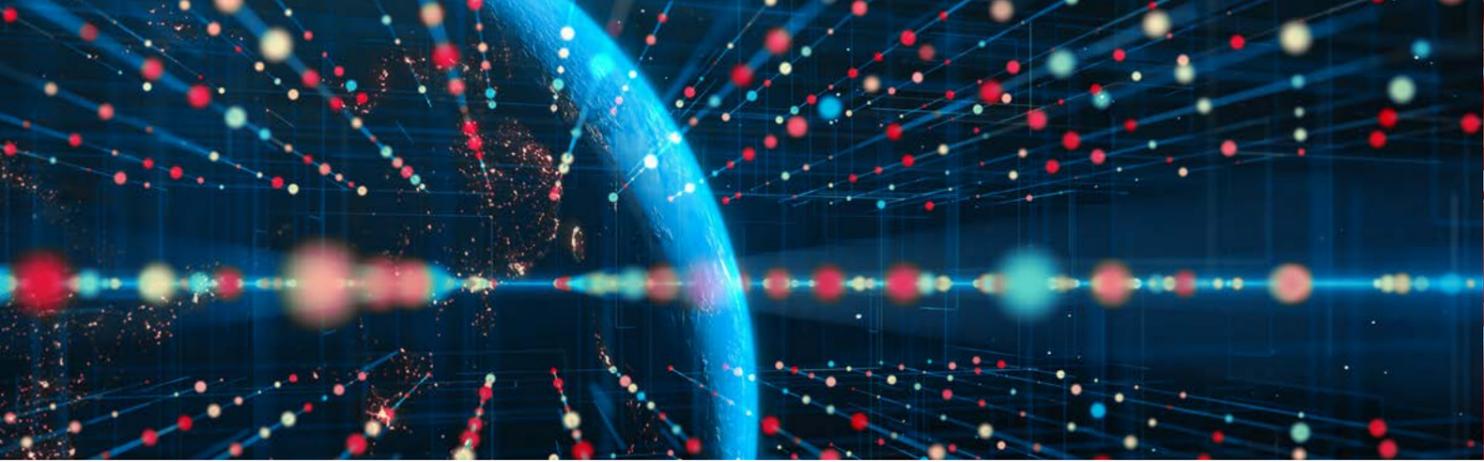
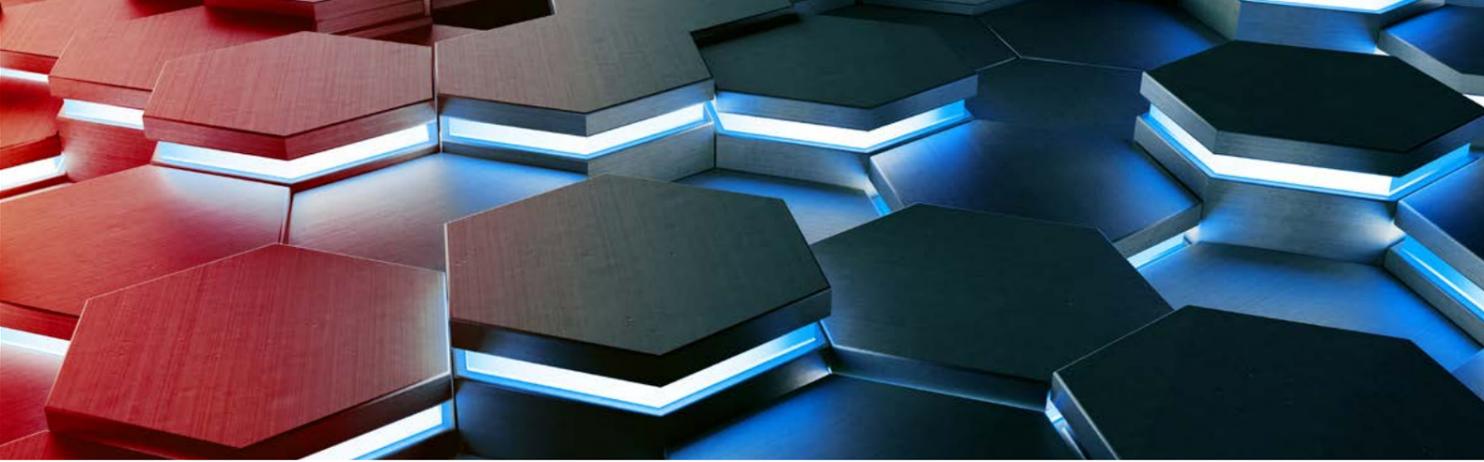

# Introduction

The discussion below documents findings from a series of workshops which were held by Thinknet 6G [1] and MÜNCHNER KREIS [2] in 2021, with the goal to provide orientation and input for developing a strategic 6G research plan. The topics selected for the workshops are aspects of 6G that we expect will have a significant impact on other industries and on society:

- 6G as both a communication infrastructure and a sensing infrastructure
- The extensive use of artificial intelligence in 6G
- The security and resilience of 6G

This paper does not go into the technical details of how to develop and implement 6G. Rather, it provides input to the wireless industry from other sectors about (mostly) non-technical topics that will need to be addressed in parallel with the technical developments, such as new use cases, regulation, communication with the public, and cross-industry cooperation.

For the sake of brevity, this paper presents a summary of the results. Please contact the authors for more detailed information about the workshops and the topics discussed. A summary of the workshops was also presented at the Thinknet 6G Summit 2021 and can be viewed on the Thinknet 6G website.[3].

### Motivation: Developing a Strategic Plan

While deployments of the 5th generation of mobile communication are still underway, research for the follow-up generation, 6G, has already begun. 6G will expand the speed and capabilities of the networks further to enable applications with significantly higher networking requirements, such as real-time digital twins or a haptic, holographic Internet. 6G moves the focus from machines to human beings and to human interaction with the physical and virtual environment.

Implementing this 6G vision over a development cycle of 10 years requires a strategic plan that defines what is needed in terms of research, technology developments, service and application enablers, standards, policies and government actions, and the building of ecosystems to create and capture value.

### Methodology

Thinknet 6G at Bayern Innovativ and MÜNCHNER KREIS jointly organized a series of three workshops focusing on 6G as the joint communication and sensing platform, to provide orientation and input for developing such a strategic plan. To ensure that input from multiple sectors was included, roughly half of the participants were from the wireless industry and the other half were from other sectors that will be important use cases for 6G, such as manufacturing, healthcare, and mobility. This provided a 360-degree view on 6G, to include opinions and issues from outside the 6G bubble.

The goal was to answer the questions:

- What are the future opportunities and threats for society and mankind that could be adressed by 6G in the 2030s?
- What are the services and use cases that you think would be of highest value for you in the 2030s?
- What are the key indicators of value and performance to describe the business impact you would want to see from 6G in the 2030s?
- What do you expect from governments and regulation in the context of 6G?
- What should be the key topics of 6G research?

The three workshops focused on the topics:

- The 6G Network as a Multi-Sensor
- AI/ML -Native Communication and Network Adaption
- Security, Privacy, Trust and Resilience in 6G

These three domains were selected because they were considered to be topics that will have significant impact on other industries and on society as a whole.

# Six Insights into 6G

Meeting the vision for 6G involves much more than research and development of new technologies. There are multiple non-technical factors that will be critical to the success of 6G and which need to be considered from the very beginning. The six insights listed below are based on groups of topics that arose during the workshops and have been confirmed not only by experts from the wireless industry but also by experts from other sectors.

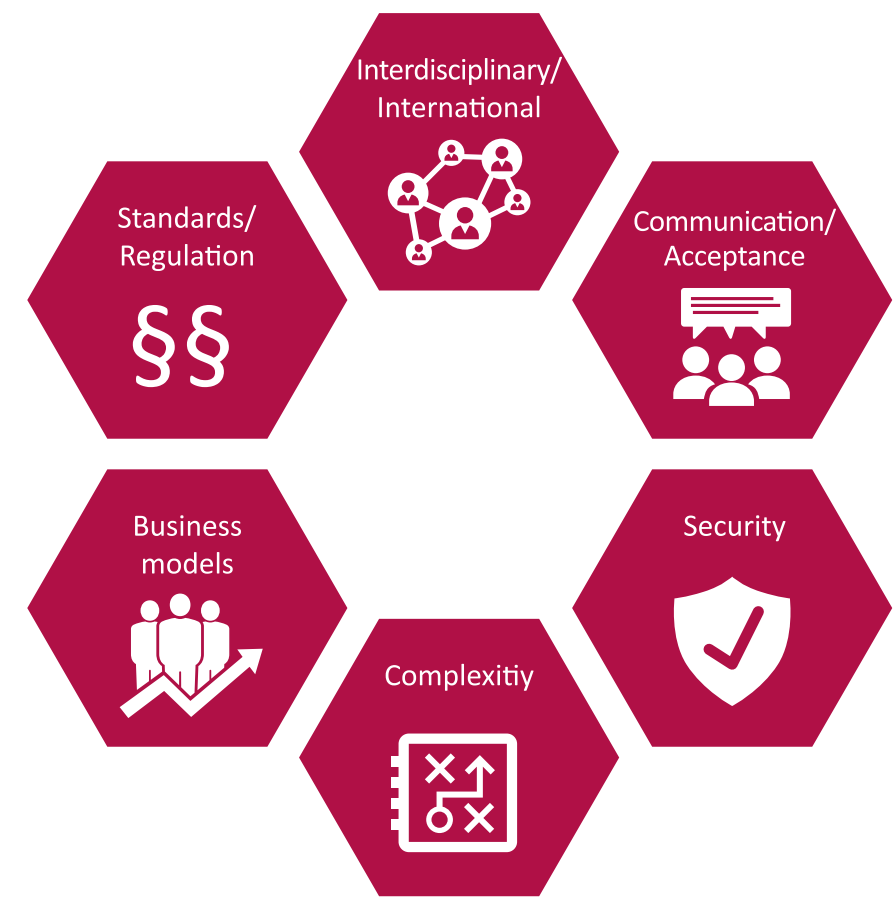

Figure 1: Six areas with significant impact for 6G

### 1. Cross-sectoral, Interdisciplinary and International Research and Development

The full potential of 6G can only be achieved when different industries and domains communicate and cooperate with each other to utilize the power of 6G. For example, end-to-end optimization and end-to-end security can only be reached through joint collaboration across multiple stakeholders.

Industry sectors that will be the main use cases for the sensing infrastructure must be included in the discussion from an early stage, to ensure that real-world requirements are met and to ensure acceptance by those industries. These other industries use different terminology and have different development cycles, release procedures, and standardization/regulation bodies than the wireless industry. These different languages, methods and processes will need to be aligned, which will require significant effort in the communication between these industries.

Additionally, the full potential of 6G can only be reached when all stakeholders worldwide work towards the same goal. Actions must be coordinated globally to develop the standards that ensure interoperability. For example, all players worldwide would benefit from international cooperative platforms to share threat intelligence and information about vulnerabilities and fixes, thereby reducing time to fix.

The interdisciplinary communication required must also extend to regulators and governmental agencies, to help reduce two issues: that regulation is usually a bit behind the technology, and that developers don't fully comprehend the regulatory processes/restrictions.

Workshop participants from both the wireless industry and other domains stressed the difficulty in finding and retaining qualified staff. Effective 6G design and implementation requires multidisciplinary expertise and experts who understand more than one domain, e.g., who understand both security and signal processing at the same time. Cross-domain academic programs and cross-domain on-the-job training are critical to the success of 6G. Programs and initiatives to keep qualified experts in Germany and in Europe will also contribute to technological sovereignty.

Both industry and academia must establish interdisciplinary experimentation and test facilities, to understand both the possibilities and limitations for sensing in practical, "real-life" situations. Numerous areas that require additional academic and industry research were identified in the workshops. These are listed in the "Research Needs" section below.

As mentioned above, 6G will only be a success when all stakeholders worldwide work towards the same goals. However, a discussion about international cooperation inevitably involves addressing the related topics of technological sovereignty, value chains and supply chains. Technological sovereignty and an infrastructure based on European values is an important aspect for 6G for all German and European stakeholders. Research and development of 6G technologies, and then production and operation of the components, must be aligned with German and European values and legislation. All regional stakeholders have a role in increasing and ensuring sovereignty, at all levels: components, infrastructure, software, data spaces, platforms, etc.

### 2. Communication with the Public and Technology Acceptance

Issues and challenges related to public confidence in wireless technologies as well as the additional issue of confidence in AI constituted a major part of the workshop discussions.

Proactive, open, and honest communication with the public about the risks and benefits of 6G is critically important for technology acceptance. Concerns and fears in the population must be adequately addressed. Both 6G and AI are new topics and not everyone is well-informed about all facets of these technologies.

Opportunities and the roadmap must be communicated realistically, not just evangelized. Stress the benefits to society, for example the environmental benefits that can be achieved with 6G. Encourage trust in the scientific community. False information (including intentional disinformation) will need to be counteracted. When AI is involved, the resulting decisions must be understandable/explainable.

Wireless radio and security concepts are difficult for the public to understand. We need much better human-centric, human-friendly methods for communicating these concepts and decisions to the users. For example, the current terms-and-conditions for using online applications are much too complicated, demonstrating that we need better methods for humans to grant informed consent.

Technology alone can't solve the problem. It requires the involvement of people and that they understand the issues involved.

Involving professional science communicators and sociologists will facilitate understanding and then acceptance. Developing functioning, secure and easy-to-understand use-cases and success stories will demonstrate both the feasibility and value of AI for 6G.

To alleviate fears around abuse of power and monopolies, a Code of Conduct for operators, vendors, and users should be developed and standardized. To alleviate fears especially around excessive dependence on hardware components from non-European suppliers, support and fund German and European research, development, and industrial manufacturing of those components.

### 3. Security, Privacy, Trust and Resilience

Issues in IT security, data privacy, trust and resilience arose in all three workshops, not only in the workshop focused on security, underscoring the significance that security will have for the success of 6G. New concepts such as joint communication and sensing will only obtain public acceptance when they are secure, when they protect privacy, and when the public is comfortable placing trust in the systems.

Proper security in design and operation creates a unique selling point and can be a geo-political advantage or a mandatory requirement.

It is critical to embed security by design and privacy by design into all network devices and all end devices. Zero-trust networking and zero-trust environments will be needed. Trustworthy hardware platforms would provide a trust anchor that can be used as the basis for ensuring integrity of the entire system. And, due to increasing virtualization and softwarization in the wireless industry, the equivalent of trustworthy hardware platforms for virtualized environments is also needed. These trustworthy platforms will be required not only for smartphones and networking equipment, but also for low-cost IoT devices.

It was felt that current data and privacy protection rules and laws are insufficient for a unified communication and sensing infrastructure. Minimal security requirements for 6G devices must be defined and enforced, including for low-cost, low-quality sensors. However, many of the devices connecting to the wireless network have limited capacity (e.g., zero-power IoT devices) and simply don't have the computing power to perform additional security operations. Provisioning and managing a massive number of (IoT) devices, for example for security patches, will be very challenging. The 6G network will likely run in parallel with 4G and 5G for several years/decades; plan for this mixed operation, particularly with regards to security and resilience. Be prepared to handle legacy 4G/5G end devices with limited or no security, such as older IoT and consumer end devices. In addition, expect the number of reduced-capacity devices and passive IoT devices to increase, and design the security to handle both legacy and low-capacity devices.

The 6G network, especially when it is used as a joint sensing and communication infrastructure, will become even more critical to society than the current 5G network. It could be argued that resilience is the single most important aspect to consider, since high bandwidth or low latency don't matter when you don't have a connection at all. As such, it will require additional effort to ensure resilience and availability of the 6G network, such as planning for integrated 4G/5G/6G usage, for redundant infrastructure, for redundant security mechanisms, and for decentralized operating structures. Resilience and graceful degradation concepts should be designed into 6G taking into consideration not only the technological aspects (e.g., the technological causes of degradation and how technology could prevent degradation), but also for example the impact that certain graceful degradation paths might have on users, safety-critical applications, and the overall complexity of systems.

An important mechanism for increasing resilience and availability is to increase coverage and to ensure ubiquity. For example, being able to provide emergency services during a natural disaster, relies on providing enough coverage that the 6G network remains available, even if large parts of it are damaged. To improve coverage and ubiquity, the 6G sensing network will be a truly cooperative network, with direct device-to-device communication. Integration with non-terrestrial networks will provide more complete coverage, especially in remote areas or when the terrestrial network is damaged. To ensure seamless communication for devices on-the-move, the 6G network will need to be interoperable with other wireless networks, such as Wi-Fi and LoRa-WAN. This full, ubiquitous coverage is important for the democratization of digital services, especially in rural areas, and will also be fundamental for many sensing applications. However, this will also require more public infrastructure.

Developments in quantum technologies compound the security issues above. The 6G network needs to be quantum-ready to be future-proof. It must use quantum-safe cryptography, and 6G stakeholders should keep abreast of developments in quantum key distribution.

### 4. Managing Complexity, Orchestration and New Technologies

Using the 6G network for joint communication and sensing necessitates a new level of complexity not yet addressed in 5G. 6G will require fusion at multiple levels in the technology: data fusion, sensor fusion, network fusion, ... And all of this across multiple providers, network infrastructures, ICT (e.g., cloud) infrastructures, and multiple sectors and domains. Standardizing, orchestrating and maintaining control over this level of complexity will be a huge challenge.

The 6G network will need to be very dynamic and adaptable, and this flexibility must be baked into the design. Important design decisions such as how to ensure reliability and maintainability, how to implement QoS, how to reduce energy consumption, and how to manage the resulting complexity need to be discussed and agreed upon. 6G will use AI to enable end-to-end network and service optimization, providing flexible and dynamic network adaption. This optimization reduces both the amount of hardware needed and the level of human interaction needed. With efficient AI-based pre-processing at the edge (e.g., security related monitoring and data analysis) the network can filter and remove unwanted results right away. AI can also be used to enable more reliable connectivity even under difficult radio conditions. AI can be used to optimize spectrum allocation and usage, and to make sharing spectrum more efficient. However, the energy costs to train an AI can be significant and must be balanced with the expected gains from automation.

The use of AI increases the complexity involved, as gathering data and training an AI/ML model are only the first steps towards effective application of AI. After initial training, the AI must continue to learn and, if necessary, adapt its decision-making based on the additional data. However, this continuous learning has serious implications for the testing and certification processes, as well as for product liability. Active learning while the AI is in live operation includes the risk of unexpected outcomes, with potential catastrophic effects on the network reliability. Even if the learning takes place offline, continuous on-going changes will require permanent feedback loops with the developers of the AI software. And continuous on-going changes will require continuous on-going testing processes and comprehensive test facilities. Taken one step further, the continually-updated AI may even change the intended purpose of the software, leading to the requirement to continuously re-certify the software or product.

Data is needed not only for AI, but also for new cooperative networks (e.g., mesh, distributed, ad hoc). These network models can enable new use cases, but inherently require sharing data, which will require new data sharing models and protocols.

Despite the complexity involved, the 6G network must remain both secure and resilient. It is not clear how to address end-to-end security and resilience in a complex system that includes many different components and many different stakeholders.

### 5. New Business Models

Independent of the topic, stakeholders from wireless and use-case industries were unanimous that 6G technology will enable new use cases and services that the industry has not yet thought of, or that were not yet technically possible. As the technology advances and new insights are obtained, use cases will become more individualized and services will be optimized for the evolving 6G capabilities.
For example, intelligent services (e.g., dynamic pricing), higher-value context-specific services (based on deeper context insights), and better-differentiated QoS offerings become possible. We expect better user experience, not only for humans but also for IoT devices, sensors, actuators, and machines. New services, new use cases and new markets are enabled using AI. AI could even develop/create new services on-the-fly as it learns user preferences.

These new possibilities will spawn new companies and new business models which take advantage of sensing data. The interaction between the virtual and physical worlds opens up new immersive communications.

In addition, we expect new forms of cooperation between incumbent operators and interdisciplinary cooperation between enterprises from different sectors. For example, a group of operators could form a consortium with automotive OEMs, municipalities, and traffic management vendors to create a smart traffic-flow optimization application.

Openness, in all its forms, appeared in multiple discussions. Openness increases transparency, enables cross-domain collaboration, and empowers smaller companies and start-ups to participate easier in the 6G development, which will also enable new business models. Some of the "open" concepts discussed included more open source generally and more extensively, anonymized open data, open RAN (Radio Access Network), open KPIs for increased transparency into the network operations, open KPIs to improve the explainability of AI used in the network, open digital identities, and open source for edge computing.

### 6. Standardization and Regulation

6G stakeholders require the surety that regulation and standardization provide in order to develop new products. The best tools & technology can't work seamlessly together when there are no common grounds for development. In addition, international regulation and standardization provide protection for both consumers and the wireless industry.

Both regulation and standardization must be clear, must take place early in the process, and must be agreed worldwide. New in 6G is the requirement to standardize across a significantly increasing number of multiple network technologies, multiple operators, multiple sensor providers and multiple industries. Ideally, vertical industries will participate in the standardization and regulation processes and will actively incorporate their views and input, to ensure that the resulting 6G network meets their needs.

Numerous areas that require standardization and/or regulation were identified in the workshops. These are listed in the "Recommendations" section below.

For example, innovative, flexible spectrum policies will be needed, to support spectrum sharing, cognitive radio, and to support sensing opportunities. Significant challenges here include massive IoT, solving the indoor/outdoor problem, and propagation (reflection) of legacy spectrum. Early communication between researchers and regulators is important, to ensure understanding of the requirements and the need for new spectrum policies. Since spectrum allocation can take a long time, evaluate the requirements early and involve the regulators early.

As another example, we expect that some applications in the 6G sensing network will require higher availability, faster response times, and higher resilience than other applications. For example, medical or emergency services require higher KPIs than television streaming. However, providing higher prioritization for some applications may come into conflict with net neutrality, and will require societal discussions and then regulation about what to prioritize and how to prioritize.

# Recommendations

The top 10 recommendations from the series of three workshops are below. These are only an excerpt from a more complete list. For additional details, please contact the authors.

**1.**

### Interdisciplinary and intersectoral communication

All stakeholders, including industry, research, and regulators, need to communicate early, communicate clearly, and communicate transparently, with the public, with the rest of the wireless industry, with other industry sectors, with researchers and with regulators. Learn to speak the other's language. Harmonize the development cycles, release procedures, standardization, and regulation across the relevant sectors as much as possible, and define standardized interfaces for exchange between industries.

**2.**

### International cooperation

Cooperate early and internationally, with other stakeholders worldwide. Coordinate activities internationally and work together towards global standards for 6G and AI. Build and foster innovative ecosystems to explore 6G use cases and to co-create applications and services that leverage 6G networks.

**3.**

### Develop and enforce standards and regulation

Develop and enforce standardization and regulation, not only for the 6G network, but also for the related supporting technologies such as collection and use of data and the use of AI. User requirements specific to certain industry sectors must also be taken into account. Additionally, regulation should regularly assess the risks that 6G is exposed to and should ensure that risk-based approaches to 6G security are covering at least these risks.

Develop standards, regulatory frameworks, and legal frameworks for:
- Enforcing coverage obligations, especially in rural areas
- Spectrum allocation and usage, including differentiated requirements for spectrum (e.g., licensed vs. unlicensed vs. lightly licensed)
- Privacy and data security
- Data collection and usage
- The use of AI in communication and sensing networks
- Security and reliability, including mandatory minimal security in all end devices
- Security conformance testing
- Dealing with errors and failures, as well as with risks that cannot yet be foreseen
- Dealing with security vulnerabilities and cyber-attacks, and liability for zero-day attacks
- Regulation and protection for white-hat hackers
- Openness (open RAN, open APIs, open data sets, open digital identities, ...)
- Regulation for new 6G campus networks (analogous to current 5G campus networks)
- EU-wide regulation and standardization for interfaces and security software architectures which are based on technology-open, hardware-independent approaches, to remain competitive in a global market

The testability and certification of both AI-based systems and security systems is a significant challenge for standardization and regulation bodies. Security requirements change constantly as new threats arise, so industry needs a dynamic approach in which they constantly adapt their systems. Certification and regulation must consider and enable dynamic changes.

**4.**

### Trust and communication with the public

Maintain proactive, ongoing communication activities to ensure open and honest discourse with the public, covering both the risks and benefits of 6G. Address the concerns and fears in the population adequately, while stressing the benefits to society. Encourage trust in the scientific community and counteract false information. Involve professional science communicators and sociologists for optimal results.

**5.**

### Invest in building up and retaining expertise

Contribute to technological sovereignty by building up skilled experts and by establishing testing facilities. Invest in education and training to build up multi-domain expertise, such as in 6G and smart manufacturing, 6G and AI, or 6G and IT security. Invest in training, re-training and retaining experts in Germany and in the EU.
Establish experimentation and test facilities, not only for academic research but also for industry use. Use these facilities to understand both the possibilities and limitations for sensing in practical, "real-life" situations.

**6.**

### Security-by-design and privacy-by-design

Ensure that security, privacy, trust, safety, and resilience are baked into the design from day one. Make security an integral part of the 6G network, not an add-on. Emphasize security-by-design and privacy-by-design from the very beginning. Ensure that 6G networks use quantum-safe cryptography Address the market failure of cybersecurity by incentivizing and ensuring enough investments into the security of products; this needs to be a joint effort of regulation, vendors, and buyers of secure products and services.
To alleviate fears around hardware components from non-European suppliers, support German and European research, development, and industrial manufacturing.

**7.**

### Focus on explainable, reliable AI with measurable KPIs

To help mitigate risks due to AI, develop quality indicators and standards for AI-based systems. In addition, there must be emergency mechanisms to prevent serious issues should the AI fail or run out of control, and there must be adequate safeguards against "hype amplification" algorithms.
To alleviate fears around abuse of power and monopolies, an AI Code of Conduct for operators, vendors, and users should be developed and standardized.
Develop KPIs for AI/ML, with a focus on quantifying AI's contribution to the network, to ensure that AI/ML's contribution to 6G services is measurable/quantifiable.

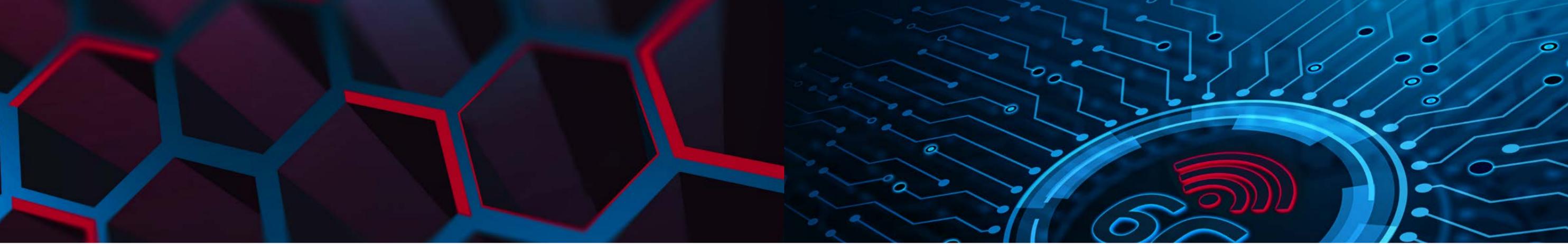

## Research Needs

The top 10 research needs identified in the workshops are listed below, in no particular order of importance or priority. In all workshops, it was emphasized that these key research areas require multidisciplinary research and a more holistic system-level view. Also, it became very apparent that many of these focus areas are closely interrelated, calling for research collaboration across disciplines such as networking, AI, data technologies, and security. Both industry and academia must establish interdisciplinary experimentation and test facilities, to understand both the possibilities and limitations of the research areas listed below.

### 1. Joint communication and sensing

The combination of sensing and communication will be a key feature of 6G. Radio system research needs to investigate, for example, how integrated communication and sensing can efficiently utilize shared spectrum and hardware resources, and how usage of these resources needs to be coordinated in order to achieve optimal performance for both sensing and communication, in various usage scenarios including in- and out-door. AI will be a key enabling technology for the 6G air interface, relying, for example, on intelligent reflecting surfaces. Resulting research questions will include which ML models and algorithms are most appropriate, what kinds of data are needed, and how the data can be provided and accessed in a secure and privacy-protecting way. Further research questions are related to the operation of such a joint communication and sensing infrastructure across multiple network operator domains and how the sensing capabilities can be used to reduce power consumption, to achieve high-precision sensing, and to enable applications such as advanced mixed reality.

### 2. 6G IoT

In order to extend the network support for highly demanding IoT settings beyond what 5G is offering, and to provide more tailored, personalized and contextually-located connectivity for different kinds of devices (wearables, augmented reality devices, etc.) and application scenarios (mission- and safety-critical industrial IoT, digital twins, etc.), research is needed across multiple areas including radio, network and cloud, AI, and data, as well as security and resilience. Main research targets will be the integration of the 5G capabilities of massive Machine-Type Communication (mMTC) and Ultra Reliable Low Latency Communication (URLLC) into the 6G capability of massive URLLC (mURLLC), including the exploration of THz bands, the support of passive IoT devices and devices with reduced capabilities (RedCAP), AI-enabled seamless and reliable network coverage, flexible function splits between devices and network through data processing in the edge, and enhanced IoT security, for example, based on sensor identities.

### 8. Form data alliances and common data pools

Cooperate with stakeholders from other industries to establish data alliances and shared data pools that incorporate data from both the sector-specific sensors and the 6G network. This will enable all stakeholders in an ecosystem to enrich their applications with additional data from other sources, thereby enabling cross-sectoral applications. For example, an application that monitors reforestation efforts could combine data from satellite images of the area, data about the number of seedlings planted in previous years, and data about weather patterns to create a multi-dimensional analysis of the success of the reforestation program. But this is only possible if this application has access to all three data sources, and access to data in an appropriate format. Develop standardized data pools and data alliances for AI and ML, to share data for AI training, active learning, and continuous learning. This will also involve defining common data formats and AI models, and could include the exchange not only of data but also of pre-trained models.

### 9. Managing system complexity

Plan and design for extreme complexity, due to new, dynamic network topologies, data fusion, sensor fusion, network fusion and platform fusion, across multiple heterogenous providers and infrastructures. Intensify efforts for a holistic system engineering, apply system level simulations to get insights into the complex behavior of future 6G systems, and consider approaches developed by complexity sciences.

### 10. Focus on resilience and availability

The 6G network, especially when it is used as a joint communication and sensing infrastructure, will become even more critical to society than the current 5G network. Ensure universal availability and extreme resilience of the network by enforcing coverage obligations and through rigorous design and planning, including planning for integrated 4G/5G/6G usage, redundant infrastructure, redundant security mechanisms, support for legacy and reduced-capacity end devices, local energy storage concepts, integration with non-terrestrial networks, and decentralized operating structures.

### 3. 6G end-to-end networking

6G will require seamless integration and interworking with previous generations of mobile communication systems, non-terrestrial networks, and ad hoc mesh and device-to-device networks. The network integration and interworking, but also the extensive use of AI, advancements towards a cloud-based compute continuum, new deployment models, the possible impact of quantum technology, and further influencing trends and factors such as the disaggregation, sustainability and resilience of a network, will require rethinking the network architecture and topologies of 6G. Research in this demanding area will need to address the end-to-end and cross-domain optimization of the network including traffic and QoS management, de-centralized operating structures as well as issues around network neutrality.

As part of future-proofing the network, 6G stakeholders should advance research in quantum technologies relevant to communication networks, such as quantum communication based on entanglement and quantum-enhanced network emulation, as potential networking resources.

### 4. Resilient complex systems

There is the risk that the growing complexity of 6G systems will have a negative impact on the reliability, availability, security, and maintainability of 6G. System research and complexity science are expected to explore approaches and solutions for engineering and managing this complexity, and to make 6G more resilient in the face of system failures and service disruptions. Resilience could be achieved through the intelligent use of redundant network resources available thanks to the integration of terrestrial and non-terrestrial networks (see above "6G end-to-end networking"), through automated, autonomous, and adaptive network management mechanisms based on AI and the capabilities of a cognitive cloud infrastructure, and through digital network twins that allow testing and predicting the system behavior under stress conditions.

### 5. Service enablement

6G will support a very broad range of applications across various sectors. Use-case-driven multi-disciplinary and cross-sectorial research will help to understand the diverse requirements and to identify common services to be provided by 6G. The expected results are services that are significantly enhanced in terms of bandwidth, latency, reliability, and energy-efficiency, but also novel services offered by the exposure of new network capabilities that will be enabled by AI and the integrated sensing. This way 6G will support green communication and the transition toward immersive communication.

### 6. Human-centricity and digital trust

Human-centricity is a design goal of 6G, making technology acceptable and usable by humans and the society. Expertise from the social sciences and the humanities is required to understand what it means to put the human into the center of the design and operation of 6G. User experience, usability, trustworthiness, accessibility, and the avoidance of harm from use are main aspects that need to be considered. Objectives and expected results in this research area include the support for context-aware personalized services, transparency and domain-specific trust models and frameworks, the support for detecting disinformation, personal data sovereignty using for example smart contracts, safety and security in human-machine interaction, the democratization of technology access, and platform and service models that are open, fair and reduce the risk of a monopoly of a few global players.

### 7. Security and privacy

Both security and privacy need to be built into the design of 6G. They require a holistic approach comprising technology, processes, and people. Research must address all three aspects and needs to investigate how to design and implement a multi-layer zero-trust architecture that performs continuous security and health checks and adapts to emerging threats. Understanding the human factor, developing and running awareness programs, and designing security and privacy in a way that supports ease-of-use and provides transparency on current security levels are research issues related to the human aspect. Research targets addressing the process dimension include the assurance of compliance with security and privacy regulations and policies, security certifications, and the sharing of threat intelligence. Technology research needs to cover jamming protection, 6G honeypots, digital identities and how to prevent identity theft, quantum safe cryptography and quantum key distribution, privacy-preserving technologies including homomorphic encryption, AI-enabled security mechanisms, but also the detection and mitigation of vulnerabilities of AI and ML models. Research will need to include the role of digital network twins for simulating cyber-attacks, but also understanding the new vulnerabilities that digital twin concepts might introduce. Specific issues that are still subject to research are IoT security and the impact that cybersecurity has on safety, new approaches for security verification and security testing using, for example, chaos engineering, how to manage the rollout of security fixes to a massive number of devices, and how to design a secure system built upon elements that cannot be trusted.

### 8. AI and ML for wireless networking

Understanding how to apply AI/ML at all layers of a wireless communication system will require interdisciplinary research including close cooperation between AI experts and domain-specific research teams. This includes not only research into multi-domain data sharing and data sharing protocols, but also multi-domain AI models, how to share pre-trained models, and enabling active continual learning. Active learning while the AI is in live operation includes the risk of unexpected outcomes, with potential catastrophic effects on the network reliability, which can be mitigated by improvements in handling catastrophic interference phenomenon. Research into recognizing and preventing misuse of AI and failure of AI will be critical to the reliability and resilience of the 6G network.

In addition to learning to apply AI to wireless communications, research in the areas of explainable AI, traceable AI and the ethics of AI is needed, to increase trustworthiness and acceptance of AI technology by the public and by the wireless industry. Emerging trends in AI such as causal AI must also be investigated and (when appropriate) applied to AI for wireless communications. Reducing energy consumption by AI, e.g., for AI training, is a vital issue for all AI applications, including AI/ML for 6G.

### 9. Data

When the 6G network is used as a joint communication and sensing platform, this sensing will generate large amounts of valuable data. In addition, AI/ML will be used extensively in 6G and will rely heavily on high-quality data. The data from sensing and for AI/ML raises significant open questions about how to unify, interpret, translate, and annotate data from multiple sources and from multiple verticals, because a lack of interoperability will impede gaining insight from the data. There are additional questions of how to ensure privacy and data security and how to define and enforce data ownership.

Depending on the applications envisioned, it may be necessary not only to collect data, but also to recognize sentiment or intent from the data.

High data quality is critical, so recognizing and preventing bias in the datasets is an important area of research. Included in data quality are research questions about how to ensure data completeness and how to recognize when data is missing, as well as how to discern whether outliers are simply measurement errors or important signals in the data.

To improve not only data quality but also to increase trust in the 6G network, methods to identify and mitigate against fake data are needed.

### 10. Softwarization

Major IT trends, including SDN (Software Defined Networking), NFV (Network Function Virtualization), virtualization, and cloud and edge computing have and will continue to affect the wireless communications industry. At the same time, new developments in specialized hardware (e.g. hardware accelerators for AI) driven by e.g.energy-efficiency will require an abstraction level above this hardware, and standardized programming interfaces to these optimized hardware components.
Functionality relevant to the 6G network will be found across the entire compute continuum, including edge, fog, core, cloud, corporate data centers, and end devices. Areas for research in softwarization include methods to extend software and virtualization to all domains of the 6G network, and software engineering methods to partition, modularize, develop, verify, deploy and orchestrate multiple smaller, distributed software components, as well as new APIs and protocols for communication and synchronization between the software components.

Softwarization enables entirely new network architectures, including the possibility to change the network architecture dynamically during operation. Research into optimal network architectures and methods to dynamically adjust the architecture (possibly using AI) during live operation are new research areas related to wireless communication. The move to softwarization is a benefit for 6G researchers, who can test innovative new ideas much more easily with software than with hardware.

As the importance of software increases, so too the importance of software dependability/resilience, security, scalability, efficiency, and performance. Research into low-code and no-code platforms to generate software which is compliant with security/privacy regulations "by design", as well as how best to utilize such platforms in 6G, is needed. The role of microkernels to improve total system security should be investigated. Research into formal verification of software, especially components related to security, will be required, as well as development methods for writing formally-verifiable software.

Efficiency and performance aspects, as well as methods to reduce energy consumption in the 6G network, include improving the performance of code generated by no-code and low-code platforms, improvements in algorithm design, advances in "green coding" (e.g. the use of more energy-efficient programming languages) and other ways to reduce energy needs in the network (e.g. via more efficient communication protocols) are needed.

# Summary and Next Steps

Implementing the 6G vision over a development cycle of 10 years requires a strategic plan that defines what is needed in terms of research, technology developments, service and application enablers, standards, policies and government actions, and the building of ecosystems to create and capture value. The goal of the workshop series organized by Thinknet 6G and MÜNCHNER KREIS and the goal of this paper is to provide orientation and input for companies and institutions to develop such a strategic plan.

Since 6G is expected to hit the market around 2030, companies and research institutions need to begin their work on 6G without delay. As the insights and recommendations detailed above illustrate, there are multiple areas where significant research and development for 6G is needed. Every stakeholder organization will need to develop an internal plan for their corporate or academic research and development. The workshop results described in this paper provide input for a more specific corporate or research 6G roadmap.

As the insights and recommendations listed above also illustrate, collaboration and interdisciplinary cooperation will be critical to the success of 6G. Both Thinknet 6G and MÜNCHNER KREIS invite interested 6G stakeholders to reach out with feedback and to participate in our think tanks and networks.

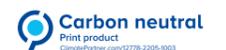

When it comes to innovations, Bayern Innovativ is knowledge manager, idea generator and catalyst. As a neutral institution of the Free State of Bavaria, Bayern Innovativ pools relevant expert knowledge especially for small and medium-sized enterprises, so they can successfully implement their innovations. Bayern Innovativ supports the innovation process as the heart and backbone of an open and sustainable ecosystem that is fit for the future. To make innovation possible, we connect people with a wide range of knowledge from industry, science, administration and politics. We manage, network and drive innovation projects — with a lot of experience and state-of-the-art methods. For our digital platforms and for connecting our clients, we are already making use of artificial intelligence today. Our vision is a Bavaria in which every viable idea becomes innovation. We turn Bavarian companies into the technological winners of tomorrow and strengthen Bavaria's business location sustainably.
**www.bayern-innovativ.de**

Thinknet 6G, hosted at Bayern Innovativ, is a think tank and network community for all organizations and stakeholders worldwide who are interested in any aspect of 6G. Thinknet 6G stimulates networking and exchange within the wireless industry, and between the wireless industry and other sectors who are important use cases for 6G. We invite all 6G stakeholders to participate in our Thinknet 6G activities and to join us at our flagship event, the Thinknet 6G Summit.
**www.thinknet-6g.com | www.thinknet-6g-summit.com**

The MÜNCHNER KREIS is the leading independent platform providing orientation for decision makers in the digital world. As a nonprofit association, the MÜNCHNER KREIS serves as an independent, interdisciplinary, and international platform for active and diverse discussions amongst key players from business, academia, and public policy. In our various activities, we analyze future developments, to provide valuable impulses on the technical, economic, political, and social challenges of the digital transformation.
**www.muenchner-kreis.de**

**AUTHORS**

Kimberley Trommler
Head of Thinknet 6G
Bayern Innovativ GmbH
trommler@bayern-innovativ.de

Matthias Hafner
Project Manager, Thinknet 6G
Bayern Innovativ GmbH
matthias.hafner@bayern-innovativ.de

Prof. Dr. Wolfgang Kellerer
Research Chair, Thinknet 6G
wolfgang.kellerer@tum.de

Peter Merz
Industry Chair, Thinknet 6G
peter.merz@nokia.com

Sigurd Schuster
Head of Working Group for Digital Infrastructure and Services,
MÜNCHNER KREIS
sigurd.schuster.ext@nokia.com

Josef Urban
Member of the Research Committee,
MÜNCHNER KREIS
josef.urban@nokia-bell-labs.com

Uwe Bäder
Advisor to the Workshop Committee
Uwe.Baeder@rohde-schwarz.com

Dr. Bertram Gunzelmann
Advisor to the Workshop Committee
bgunzelmann@apple.com

Andreas Kornbichler
Advisor to the Workshop Committee
andreas.kornbichler@siemens.com

22045 werbersbuero.de

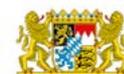